\documentclass[11pt]{elsarticle}
\usepackage{graphicx,color,amsmath,amssymb}
\usepackage{bm}
\usepackage[colorlinks=true, pdfstartview=FitV, linkcolor=red, citecolor=blue, urlcolor=blue]{hyperref}
\usepackage{ulem}

\usepackage{lipsum}

\makeatletter
\def\ps@pprintTitle{%
 \let\@oddhead\@empty
 \let\@evenhead\@empty
 \def\@oddfoot{}%
 \let\@evenfoot\@oddfoot}
\makeatother
\begin{document}

\begin{frontmatter}
\title{The Glasma, Photons and the Implications of Anisotropy}

\author[bnl,rbrc,ccnu] {Larry McLerran}
\ead{mclerran@bnl.gov}
\author[bnl]{Bj\"orn Schenke}
\ead{bschenke@bnl.gov}

\address[bnl]{Physics Dept, Bldg. 510A, Brookhaven National Laboratory, Upton, NY-11973, USA}
\address[rbrc]{RIKEN BNL Research Center, Brookhaven National Laboratory, Upton NY 11973, USA}
\address[ccnu]{Physics Dept, China Central Normal University, Wuhan, China}

\begin{abstract}

We introduce distribution functions for quarks and gluons in the Glasma and discuss how they satisfy various relationships of statistical physics.
We use these distributions to compute photon production in the early stages of heavy ion collisions.  
Photon rates satisfy geometric scaling, that is, the emission rate per unit area scales as a function of the saturation momenta
divided by the transverse momentum of the photon.  
Photon distributions from the Glasma are steeper than those computed in the Thermalized Quark Gluon Plasma (TQGP).
Both the delayed equilibration of the Glasma and a possible anisotropy in the pressure lead to slower expansion and
mean times of photon emission of fixed energy are increased.
This delayed emission might allow for larger photon elliptic flow.

\end{abstract}
\end{frontmatter}

\section{Introduction}

There is a discrepancy between what is experimentally observed in heavy ion collisions \cite{Adare:2011zr,Lohner:2012ct}, and
theoretical computations of photon emission from a thermalized 
Quark Gluon Plasma (TQGP) \cite{Shen:2013cca,Shen:2013vja,vanHees:2011vb,Bratkovskaya:2008iq,Linnyk:2013hta}.
Computations using realistic equations of state and state of the art computations of photon emission
from the TQGP give results whose slope in $p_T$ is too shallow (falls too slowly in $p_T$) and is about a factor of 4 smaller than the observed rate  
\cite{Shen:2013vja}.

Even more significant is the difference between theoretically computed and experimentally observed photon elliptic flow.  Photon emission is dominated by early times when temperatures are high and flow is not yet built up. This early time emission in theoretical calculations leads to a significant under-estimation of elliptic flow compared to experimental data.
This has led to suggestions that the photon rate may be due to late time hadronic processes where the spectrum of photons is enhanced by radial flow effects \cite{Shen:2013cca,Shen:2013vja,vanHees:2011vb,Bratkovskaya:2008iq,Linnyk:2013hta}.  This is perhaps possible but difficult, as it involves rates for photon production that are quite large, and for $3-4\,{\rm GeV}$ photons requires substantial Lorentz boosts of fundamental emission processes. 

On the other hand, a recent analysis shows that the photon spectrum measured at RHIC and LHC for p+p, d+Au, Au+Au collisions at RHIC energy
and Pb+Pb collisions at LHC is well described by the geometric scaling hypothesis \cite{Klein-Bosing:2014uaa}.  This follows from theories of gluon saturation, and should be preserved when the expansion is scale invariant, suggesting that the photons are emitted early in time when particle masses are not important.  However, one faces the problem of photon flow, an effect which takes a time of the order the transverse size of the colliding region to set in.

In this paper we suggest that a description of the photon spectrum, in particular the photon flow, should be based on the picture that the 
early dynamics of the collision are described by a Glasma, not a thermalized Quark Gluon Plasma.  
Photon emission from the Glasma was considered in a previous work \cite{Chiu:2012ij}. Following recent developments \cite{Gelis:2013rba}, we will assume that the Glasma is described by hydrodynamic equations with perhaps some asymmetry between longitudinal and transverse pressure \cite{Blaizot:2011xf, Kurkela:2011ti}.
We fully understand that the assumption that the Glasma expands hydrodynamically with near perfect fluid behavior is contentious, and as well
not much is known from first principles concerning the distributions of gluons in the Glasma \cite{Berges:2013eia,Berges:2013fga,Berges:2013lsa}.
We will nevertheless proceed assuming that such near perfect fluid behavior is true, and use the simplest possible model, developed in this paper,
to describe the quarks and gluon distributions in the Glasma.  Hopefully with deeper understanding the semi-quantitative conclusions found here survive
a much more sophisticated treatment.

Hadron distributions are generated at late times during decoupling, and are affected little by the Glasma, except as a remnant of hydrodynamic expansion.
Photon  emission is however different. In addition to knowledge of the bulk dynamics, one needs to know the quark and gluon distribution functions
as a function of the energy density of the system at early and intermediate times in order to compute the distribution of emitted photons. 
We argue that these distribution functions and their time evolution can be quite different between the TQGP and the Glasma.

A qualitative effect we find is related to the two scales that characterize gluon distributions in the Glasma.  There is an ultraviolet scale $\Lambda$ above which distributions become dilute and an infrared scale $\Lambda_{\rm IR}$ at which the distributions are highly coherent.  For a thermal system,
the UV scale is the temperature, and the  IR scale is the magnetic has $M_{\rm mag} \sim \alpha_s T$.  In the Glasma at the initial time,
both the infrared and UV scale are equal and are proportional to the saturation momentum.  (Within our analysis, we  determine
the numerical relation between these initial scales and the saturation momentum.)  As time evolves these scales split apart.  Thermalization can begin when these scales reach $\Lambda_{\rm IR}(t) \sim \alpha_s \Lambda(t)$ \cite{Blaizot:2011xf, Kurkela:2011ti}.  (In the later analysis we will make this criterion more precise.)  For a thermal system, the temperature scale decreases approximately as $1/t^{1/3}$. However, for the Glasma, we will show below that the infrared scale decreases more rapidly than this but the UV scale decreases more slowly.  Such evolution is slowed down even more if there is an asymmetry between longitudinal and transverse pressure.

Therefore, if we start both the Glasma and the TQGP from the same initial ultraviolet energy scale,  it will take longer in the Glasma to reach the same ultraviolet energy scale at a lower value. We will show that this has several consequences:
The shape of the photon distribution is changed and the typical time for photon emission increases.   The story for us is a bit more complicated however,
as when one properly models both the Glasma and the TQGP, the initial temperature for the TQGP must be different from that of the Glasma, in order
that they can match together at thermalization.  In addition the photons are emitted both from the Glasma at earlier times and the TQGP at later times.
We find nevertheless, that the presence of an early time Glasma slows the emission process for photons, and steepens the spectrum of produced photons.
For small asymmetry between the longitudinal and transverse pressure, we find significant steepening of the photon spectrum and significantly longer times of photon emission.
This longer emission time will allow for elliptic flow effects to become important, since flow develops on time scales of the order of the size of the system, which is parametrically large compared to the time scale set by the inverse saturation momentum.

These effects point in the correct direction but they may not be sufficient to fix the problem when these simple computations presented here 
are realized in realistic hydrodynamical or Glasma simulations. The simple considerations in this paper may nevertheless be useful in guiding such more detailed computations, where the answer to the photon puzzle can be ultimately resolved.
Of course our simple results automatically satisfy geometric scaling, and to what degree this is maintained in more detailed computations remains to be seen.   Such scaling is however very
general and should be true in ideal gas hydrodynamic models where the initial time is specified as $t_0 \sim 1/T_0$, where $T_0$ is the initial temperature.  In such a circumstance, the initial 
temperatures is determined by the multiplicity and is proportional to the saturation momentum.  The scaling follows because no dimensional scale is introduced in
the hydrodynamic expansion.

\section{A Model for Glasma Distribution Functions} 

We begin with a simple model of the Glasma evolution. We will assume the distribution function for gluons in the Glasma to be of the form
\begin{equation}\label{eq:gluons}
  g = {  {\kappa \Lambda_{\rm IR}}  \over {N_c \alpha_{S} \Lambda} }~ {1 \over {e^{E/\Lambda} - 1}}\,.
\end{equation}
In this equation, the constant $\kappa $ is of order 1.  This distribution becomes a thermal gluon distribution function when the the infrared scale satisfies
\begin{equation}
  \Lambda_{\rm IR} = {{N_c \alpha_s} \over \kappa} \Lambda\,.
\end{equation}
This is the separation of scales in a thermal system for the relation between the magnetic mass scale and the temperature.  At the magnetic mass scale 
configurations in the Quark Gluon Plasma become non-perturbative.
For $E \ll \Lambda$ this distribution is a classical thermal distribution, $T_{\rm eff}/E$, where $T_{\rm eff} = \kappa \Lambda_{\rm IR}/(N_c\alpha_s)$.  It is cut off at the scale $\Lambda$.  Before reaching the thermalization scale, this distribution represents an over-occupation of gluonic states.
For the time being, we will assume that the quarks and gluons are massless, and that the distribution functions are isotropic.

The distribution $g$ can be shown to follow from a statistical matrix $Z$, which is a generalization of the partition function.  \begin{eqnarray}
     Z & = &  \Pi_i  \sum_n { {\Gamma(\eta - n)}  \over  {\Gamma(\eta)\Gamma(n+1)} } 
e^{-nE_i/\Lambda} \nonumber \\
& = &  \Pi_i  \left( {1 \over {1-e^{-E_i/\Lambda}}} \right)^\eta \nonumber \\
& = &  e^{-\beta V \eta  \int {{d^3p} \over {(2\pi)^3}} ~\ln(1-e^{-E/\Lambda})}\,,
\end{eqnarray}
where 
\begin{equation}\label{eq:eta}
  \eta = {{\kappa \Lambda_{\rm IR}} \over {\Lambda N_c \alpha_s}}\,.
\end{equation}
The gluon distribution then follows via 
\begin{eqnarray}
     g  &  = & {1 \over Z} ~\Pi_i  \sum_n { {\Gamma(\eta - n)}  \over  {\Gamma(\eta)\Gamma(n+1)} } 
n e^{-nE_i/\Lambda} \nonumber  \\ & = & (1-e^{-E/\Lambda})^\eta  {d \over {d(-E/\Lambda)}} \left( {1 \over {1-e^{-E/\Lambda}}} \right)^\eta \nonumber \\
& = & \eta {1 \over {e^{E/\Lambda} -1} }\,.
\end{eqnarray}

For the quarks, we take
\begin{equation}\label{eq:quarks}
  q = {1 \over {e^{E/\Lambda} +1}}
\end{equation}
There is no enhancement for the quark degrees of freedom because they are fermions, as there can be no multiple occupation of fermion states.

The entropy associated with these distributions is
\begin{equation}
   S = V \int {{d^3p} \over {(2\pi)^3}} \left( (1+g)\ln(1+g) - g\ln g -(1-q)\ln(1-q) -q\ln(q) \right)\,.
\end{equation}

Using distributions (\ref{eq:gluons}) and (\ref{eq:quarks}) to express the energy density, pressures, and number densities in terms of the
quantities $\Lambda$ and $\Lambda_{\rm IR}$ we find
\begin{equation}\label{eq:edenGluons}
   \varepsilon_g = {\pi^2 \over {30}} ~2(N_c^2-1) ~{\kappa \over {N_c\alpha_s}} ~\Lambda_{\rm IR} \Lambda^3
\end{equation}
for gluons and
\begin{equation}\label{eq:edenQuarks}
    \varepsilon{q} = {\pi^2 \over {30}} ~4N_c N_f ~ {7 \over 8} ~\Lambda^4
\end{equation}
for quarks, where $N_f$ is the number of light quark flavors.
The numbers of quarks and gluons are 
\begin{equation}
  \rho_g = {{\zeta(3)} \over {\pi^2}} ~ 2(N_c^2-1)~{{\kappa} \over  {N_c \alpha_s}} \Lambda_{IR} \Lambda^2
\end{equation}
and
\begin{equation}
\rho_q= {{\zeta(3)} \over {\pi^2}} ~ 4N_c N_f ~ {3 \over 4} ~ \Lambda^3
\end{equation}

It is amusing that the entropy for the gluons is not of order $\Lambda_{\rm IR}/\alpha_s$.  The disappearance of this factor
is because the leading order term which is associated with a classical field cancels.  The classical term comes when $g$ is large, and in this limit the  
first and second terms in our expression for the entropy are both of order $g\,\ln(g)$ and they cancel. One picks up the non leading term
in the entropy density $\ln(g)$, so that the entropy is of order $\Lambda^3 \ln(\kappa/(N_c\alpha_s))$.  This means that the entropy of the 
initial state is low compared to the gluon number density, $S_g/N_g \sim \alpha_s \Lambda/\Lambda_{\rm IR}$.  At thermalization,
when $\Lambda_{\rm IR} \sim \alpha_s \Lambda$, the number density and entropy density are of the same order of magnitude.

Conformal invariance of the presented expressions guarantees that $P = E/3$ for isotropic distributions, and this is satisfied for our expressions..

We will also be interested in the case where there is a momentum an\-isotropy.  We can introduce such an anisotropy into the momentum distributions
of the particles, using that the longitudinal pressure is taken to be $P_L = \delta \varepsilon$ with $0 < \delta < 1/3$.  
This will modify the expansion dynamics through the evolution equation which follows from $1+1$ dimensional hydrodynamics, 
\begin{equation}
  \partial_t \varepsilon = {{\varepsilon + P_L} \over t}
\end{equation}
This will have the energy density fall as $\varepsilon/\varepsilon_0 \sim (t_0/t)^{1+\delta}$.

The corresponding modification of the momentum distributions can be parametrized by
$f(\mathbf{p}) = f_{\rm iso}(\sqrt{\mathbf{p}^2+\xi(\mathbf{p}\cdot \hat{\mathbf{n}})^2})$, where $\hat{\mathbf{n}}$ is the direction of
the anisotropy and $\xi$ characterizes its strength \cite{Romatschke:2003ms}.
Following \cite{Martinez:2009mf,Martinez:2009fc,Martinez:2010sc} we find the relation
\begin{equation}\label{eq:deltaXi}
  \delta = \frac{(1+\xi)\arctan(\sqrt{\xi})-\sqrt{\xi}}{\xi(1+\xi)\arctan(\sqrt{\xi})+\xi^{3/2}}\,.
\end{equation}
Note that $\lim_{\xi\rightarrow 0} \delta = 1/3$ and $\lim_{\xi\rightarrow \infty} \delta = 0$  as it should be. 
For $\xi>0$ this leads to a particle number reduced by a factor of $(1+\xi)^{-1/2}$ \cite{Romatschke:2004jh}. 
%If we require the same number of produced particles for $\delta<1/3$ as in the case $\delta=1/3$, we need to adjust the initial energy scale $\Lambda_0$.

In our computations, we will consider a fixed momentum anisotropy.  This is slightly different from what is done in standard hydrodynamic computations, where the viscosity to entropy density ratio, $\overline  \eta =\eta/s$ is taken to be a constant.  In our case, we interpret this ratio as scattering time divided by a thermal wavelength, $\sim 1/T$.  We  will argue in the next section that the scattering time for the out of equilibrium Glasma is proportional to the time,
$t_{\rm scat} \sim t$.  In this case, the expression derived from hydrodynamics becomes
\begin{equation}
  {{P_L} \over {P_T}} \sim {{3t T - 16 \overline \eta} \over {3t T + 8\overline \eta}} \sim {\rm constant}
\end{equation}

\section{Time Evolution and Parametrizing the Glasma}

The parameter $\kappa$ can be determined by knowing the ratio of the initial number of gluons to quarks in the hadron wavefunction.  At the initial time, we have $\Lambda(t_0) = \Lambda_{\rm IR}(t_0)$.  The ratio of the total number of gluons to quarks and antiquarks of all flavors is therefore
\begin{equation}
  N_{\rm gluon} /N_{\rm quark} ={4 \over 3}~ {\kappa \over {N_c \alpha_s}} {{N_c^2-1} \over {2N_cN_f}}
\end{equation}
where $N_f$ is the number of quark flavor degrees of freedom important at the saturation momentum scale.  We will take $N_f=3$ ignoring the effects of charm quarks.

In the MSTW parton distribution functions \cite{Martin:2009iq}, there is about an order of magnitude difference between the number of up quarks and the number of gluons,
so that 
\begin{equation}
  N_{\rm gluon}/N_{\rm up} = {4 \over 3} {\kappa \over {N_c \alpha_s}} {{N_c^2-1} \over N_c} \sim 10
\end{equation} 
Taking $\alpha_s \sim 0.3$, this suggests that $\kappa \sim 2-3$. 

The way that the time evolution is computed is from identifying the time with the collision time in the transport equations \cite{Blaizot:2011xf}.  Dimensional arguments give
\begin{equation}
  t \sim \Lambda /\Lambda_{\rm IR}^2 \,.
\end{equation}
We further use that the energy density scales according to the hydrodynamic equations as $\varepsilon \sim 1/t^{1+\delta}$.
The factor of $\delta$ allows for an asymmetry between longitudinal and transverse pressure as discussed above. 
Recent numerical results suggest that $\delta$ might be close to $1/3$, corresponding
to equal longitudinal and transverse pressures \cite{Gelis:2013rba}.  We will allow for some anisotropy in our analysis, but in fact find a good description of the photon spectrum and significant increase in emission times with $\delta \approx 1/3$.

Using the expression for the energy density in terms of $\Lambda$ and $\Lambda_{\rm IR}$, Eqs.\,(\ref{eq:edenGluons}) and (\ref{eq:edenQuarks}),
we can determine the evolution of these energy scales.  Note that this can also be done directly from results for $\varepsilon(t)$ obtained from hydrodynamic simulations.  We do not need to require a power law in time.

Nevertheless, the simplified structure we get when we assume a power law in time for the evolution of the energy density makes it useful to consider.

\subsection{Gluon dominated case}

Let us first assume that in the Glasma phase the evolution is dominated by the gluons and that the coupling constant is fixed.  
Then we find
\begin{equation}
\Lambda_{\rm IR} = \Lambda_0 (t_0/t)^{(4+\delta)/7}\,,
\end{equation}
and
\begin{equation}
\Lambda = \Lambda_0 (t_0/t)^{(1+2\delta)/7}\,.
\end{equation}
Note that the ultraviolet scale which corresponds roughly to a temperature falls less rapidly than is the case for a thermal system.  

For a thermally equilibrated  system
$T \sim T_0 (t_0/t)^{1/3}$, and $\Lambda_{\rm IR} \sim \alpha_s T$.

Let us consider how flow would affect a thermal system.  Let us ask what flow ensues if we are at some fixed temperature T.   For a thermal system, the initial temperature is given in terms of the particle multiplicity as
\begin{equation}
 t_0 T^3_0 \sim T_0^2 \sim {1 \over {\pi R^2}} {{dN} \over {dy}}\,.
 \end{equation} 
  Then 
\begin{equation}
  t = T_0^2/T^3 \sim \Lambda_{QCD}^2 (E / \Lambda_{QCD})^{1/4} A^{1/3} /T^3\,.
\end{equation}
where in this equation $E$ is the center of mass energy per nucleon.
From this we learn that processes at a fixed temperature will happen at a time which scales like the radius of the system and the fourth root of the beam energy.
On the other hand flow effects will become important at a time of order $A^{1/3}$.
So given a fixed temperature scale, whether or not
flow is important depends on how large is $(E^{1/4}\Lambda_{QCD}^{3/4}/T)$.  This is independent of the size of the system.  Flow effects
at some temperature scale are important or not depending upon how high the beam energy is and what value of $T$ we choose.

In the Glasma, the evolution of the ultraviolet scale is slower than the evolution of the temperature for the TQGP.
In the gluon dominated case the UV scale, for $0  < \delta < 1/3 $, cools off as $\Lambda = \Lambda_0 (t_0/t)^{(1+2\delta)/7}$, with the exponent $1/7 < (1+2\delta)/7 <  5/21$, which is always less than a third.  However, when $\Lambda_{\rm IR} = N_c \alpha_s \Lambda/\kappa$, we reach the thermalization time 
\begin{equation}
  t_{\rm th} = \left(\frac{\kappa}{N_c\alpha_s}\right)^{7/(3-\delta)} t_0\,.
\end{equation}
For the range of $\delta$ at hand, $7/3  < 7/(3-\delta) < 21/8$, and using $\alpha_s = 0.3$ and $\kappa \sim 2 - 3$, we find that thermalization may occur at a time of 1-2 orders of magnitude larger than that of the initial time scale $t_0 \sim 1/\Lambda_0$.
  This can correspond to quite large times of order $1-10\,{\rm fm}/c$. 
 
For the case of $\delta = 1/3$, the energy density of the TQGP and that of the Glasma decrease at the same rate. So when the system
thermalizes, it thermalizes at the same temperature.  However, the time it has taken to get to this scale is longer for the Glasma because it took a longer time
to evolve to this scale.   This is consistent because the initial temperature of the TQGP phase was higher than that of the Glasma.  On the other hand for $\delta < 1/3$, the situation is a bit more complicated.
Energy densities at the thermalization time must be equal, but because the Glasma is anisotropic, the average temperature will be lower in the isotropic thermal system. Also, the number of particles will not be conserved.  For the cases we consider these effects are small but they must be taken into account. 
\
\subsection{System of quarks and gluons}
Including gluons and quarks, for general $\delta$ in the Glasma case we need to solve 
\begin{equation}
  \left(\frac{t_0^{\rm Glasma}}{t}\right)^{1+\delta} = 2(N_c^2-1) \frac{8}{7} \frac{\kappa}{\alpha_s N_c} t^3 \Lambda_{\rm IR}^7 + 4 N_f N_c t^4 \Lambda_{\rm IR}^8
\end{equation}
numerically for $\Lambda_{\rm IR}(t)$. $\Lambda(t)$ follows from $t\sim \Lambda/\Lambda_{\rm IR}^2$.

The initial values $\Lambda_{\rm IR,0}=\Lambda_{\rm IR}(t_0^{\rm Glasma})$ and $\Lambda_0=\Lambda(t_0^{\rm Glasma})$ are chosen to fulfill the requirement of energy conservation
when thermalizing at the time $t_{\rm th}$, which is the time when $\eta$ (see Eq.\,(\ref{eq:eta})) is equal to $1$. 
The energy density in the thermal case is given by
\begin{equation}
  \varepsilon_{\rm th}(t) = \frac{\pi^2}{30} \left(2(N_c^2-1) + \frac{7}{2} N_c N_f\right)T(t)^4  \,,
\end{equation}
while in the Glasma we have
\begin{equation}
  \varepsilon_{\rm Glasma}(t) = \mathcal{N}(\xi) \frac{\pi^2}{30} \left(2(N_c^2-1)\frac{\kappa}{N_c \alpha_s} \Lambda_{\rm IR}(t) \Lambda(t)^3
  + \frac{7}{2} N_c N_f\Lambda(t)^4\,\right),
\end{equation}
where $\mathcal{N}(\xi) = 0.5 [1/(1+\xi)+\arctan(\sqrt{\xi})/\sqrt{\xi}]$, which results from having an anisotropic momentum distribution in the case of general $\delta$. $\xi$ can be obtained from the relation (\ref{eq:deltaXi}) for a given $\delta$.

Given an initial temperature $T_0$, the requirement 
\begin{equation}
  \varepsilon_{\rm Glasma}(t_{\rm th}) = \varepsilon_{\rm th}(t_{\rm th}) 
\end{equation}
can be fulfilled by iteratively solving for the initial $\Lambda_0=\Lambda_{\rm IR,0}$ and $t_0^{\rm Glasma}$.

The particle number in the thermal case is constant in time and given by
\begin{equation}\label{eq:dNdytherm}
  \frac{dN_{\rm th}}{dy} = \frac{2 \zeta(3)}{\pi^2} t_0 \pi R^2 \left( (N_c^2-1) +\frac{3}{2} N_c N_f\right) T_0^3\,.
\end{equation}
In the Glasma the particle number depends on time and we need to evaluate the expression 
\begin{equation}\label{eq:dNdyGlasma}
  \frac{dN_{\rm Glasma}}{dy}(t) = \frac{1}{\sqrt{\xi+1}} \frac{2 \zeta(3)}{\pi^2} t \pi R^2 \left(\frac{\kappa\Lambda_{\rm IR}(t)}{\Lambda(t)N_c\alpha_s} (N_c^2-1) +\frac{3}{2} N_c N_f\right) \Lambda(t)^3\,,
\end{equation}
at time $t=t_{\rm th}$.
For $\delta=1/3$ expressions (\ref{eq:dNdytherm})  and (\ref{eq:dNdyGlasma}) agree at time $t_{\rm th}$. For general $\delta$ there will be a difference due to the anisotropic momentum distributions in the Glasma expressions.

\begin{figure}[h]
\begin{center}
\includegraphics[width=12cm]{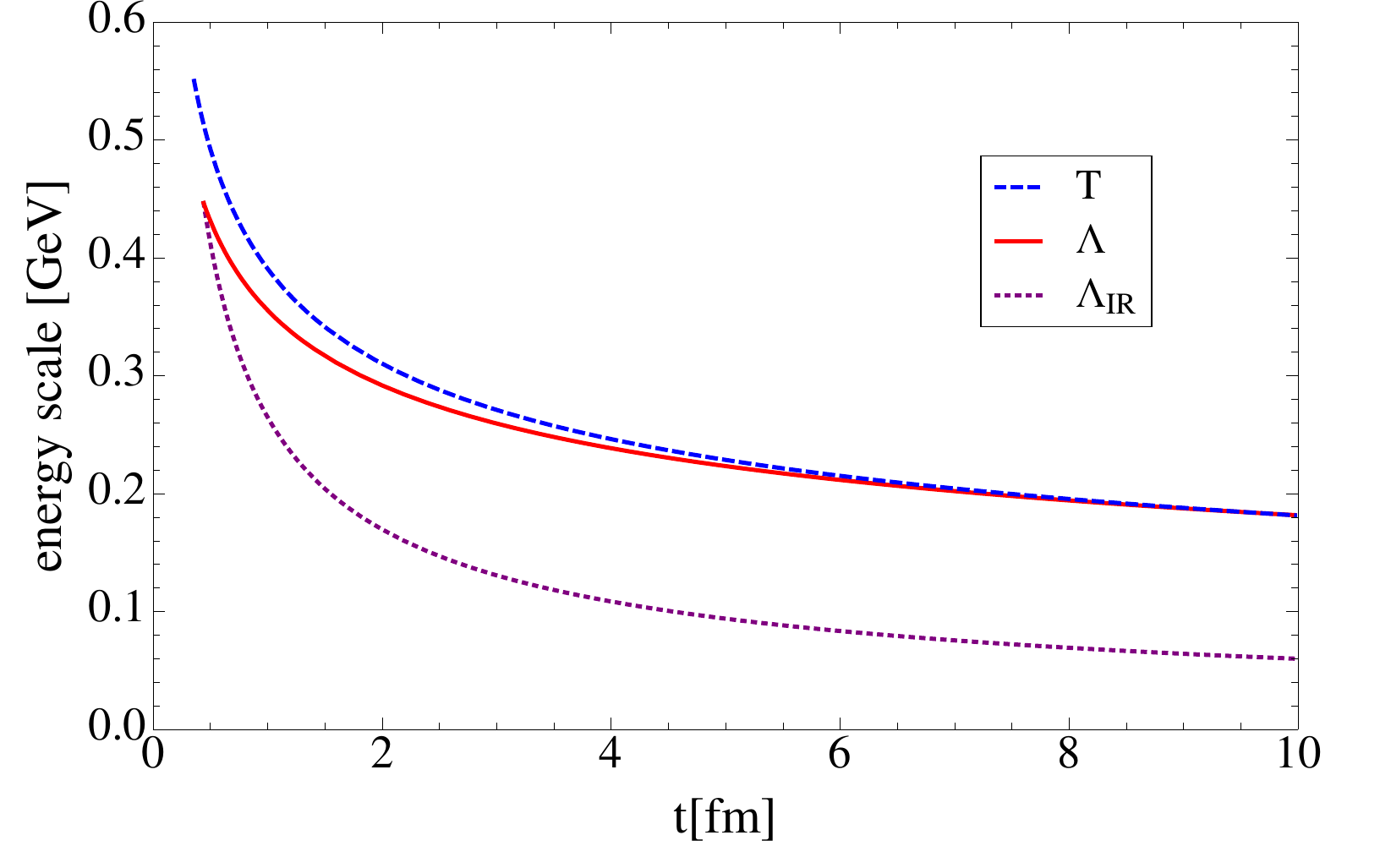}
\caption{Time evolution of the temperature $T$ in a thermal system and the ultraviolet scale $\Lambda$ and infrared scale $\Lambda_{\rm IR}$ in the Glasma for $\delta=1/3$. \label{fig:energyScales-delta1over3}}
\end{center} 
\end{figure}

\begin{figure}[h]
\begin{center}
\includegraphics[width=12cm]{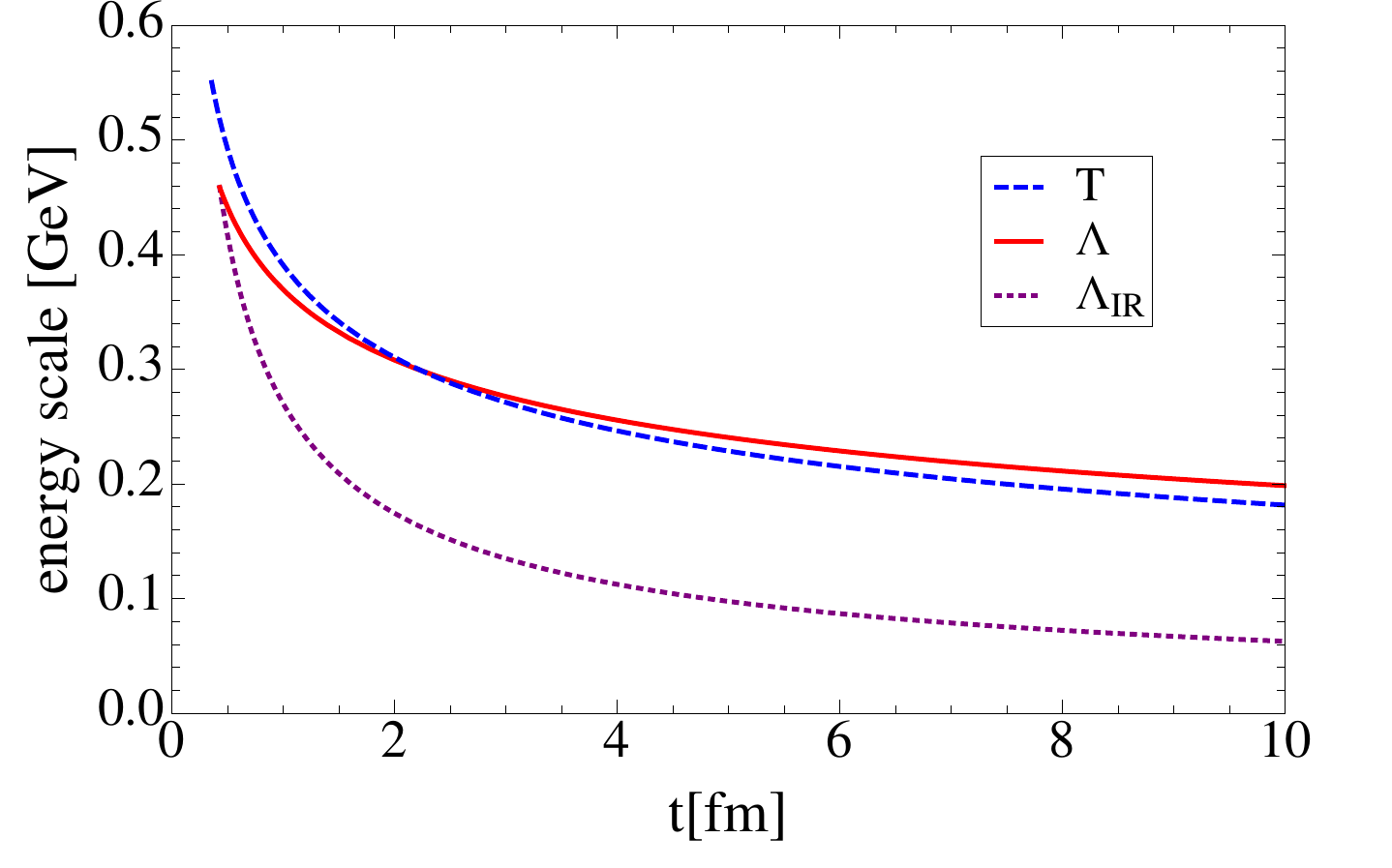}
\caption{Time evolution of the temperature $T$ in a thermal system and the ultraviolet scale $\Lambda$ and infrared scale $\Lambda_{\rm IR}$ in the Glasma for $\delta=1/4$. \label{fig:energyScales-delta1over4}}
\end{center} 
\end{figure}

For collisions at LHC energies we take $T_0=550\,{\rm MeV}$, $\alpha_s=0.3$, $N_c=3$, $\kappa=3$.
Figs.\,\ref{fig:energyScales-delta1over3} and \ref{fig:energyScales-delta1over4} show the resulting time evolution of $T$, $\Lambda$, and $\Lambda_{\rm IR}$
for $\delta=1/3$ and $\delta=1/4$, respectively.

For  $\delta=1/3$ the thermalization time is $\sim 9.7\,{\rm fm}$. In this case the temperature and multiplicity agree exactly with the thermal system at the time of thermalization. For $\delta=1/4$ the thermalization time is $\sim 8.6\,{\rm fm}$. As one can see in Fig.\,\ref{fig:energyScales-delta1over4}, the ultraviolet scales are not exactly the same at this time, which is caused by the anisotropic momentum distribution in the glasma case. In a similar way the particle number in the Glasma at thermalization differs from that in the thermal system by approximately $1\%$.

\section{Photon production from the Glasma and the thermalized QGP}
The leading order photon production rates from Compton and annihilation processes in a thermalized QGP are given by \cite{Kapusta:1991qp}
\begin{equation}
  E \frac{dN_{\rm th}}{d^4x d^3p} = \frac{5}{9}{{\alpha_s \alpha} \over {2\pi^2}} T^2 e^{-E/T} \ln\left(\frac{2.912}{4\pi\alpha_s}\frac{E}{T}\right)\,.
\end{equation}
In the Glasma we use the generalized expression
\begin{equation}\label{eq:GlasmaRate}
  E \frac{dN_{\rm Glasma}}{d^4x d^3p} = \frac{5}{9}{{ \alpha} \over {2\pi^2}} {\kappa \over {N_c\alpha_s}} \Lambda_{\rm IR} \Lambda e^{-E/\Lambda}\ln\left(2.912\frac{E}{\Lambda_{\rm IR}} \frac{N_c}{4\pi\kappa}\right)\,.
\end{equation}
Eq.\,(\ref{eq:GlasmaRate}) is an approximation ignoring the anisotropy in the momentum space distribution of the quarks and gluons when $\delta\neq 1/3$. The generalized rates taking this anisotropy into account have been computed in \cite{Schenke:2006yp}. Because here we are focusing on the effect of the energy scales and not on the rapidity dependence of photon production this approximation is appropriate.
We further use the fact that the gluon distribution in the initial or final state is large compared to 1. This leads to the simple form (\ref{eq:GlasmaRate}) where we get an additional factor $\kappa \Lambda_{\rm IR}/(N_c\alpha_s \Lambda)$ from the gluon distribution.
We chose the factor under the logarithm in the Glasma rate as to make the rates equal at the time of thermalization when $\Lambda_{\rm IR} = \Lambda N_c \alpha_s /\kappa$.

\begin{figure}[h]
\begin{center}
\includegraphics[width=12cm]{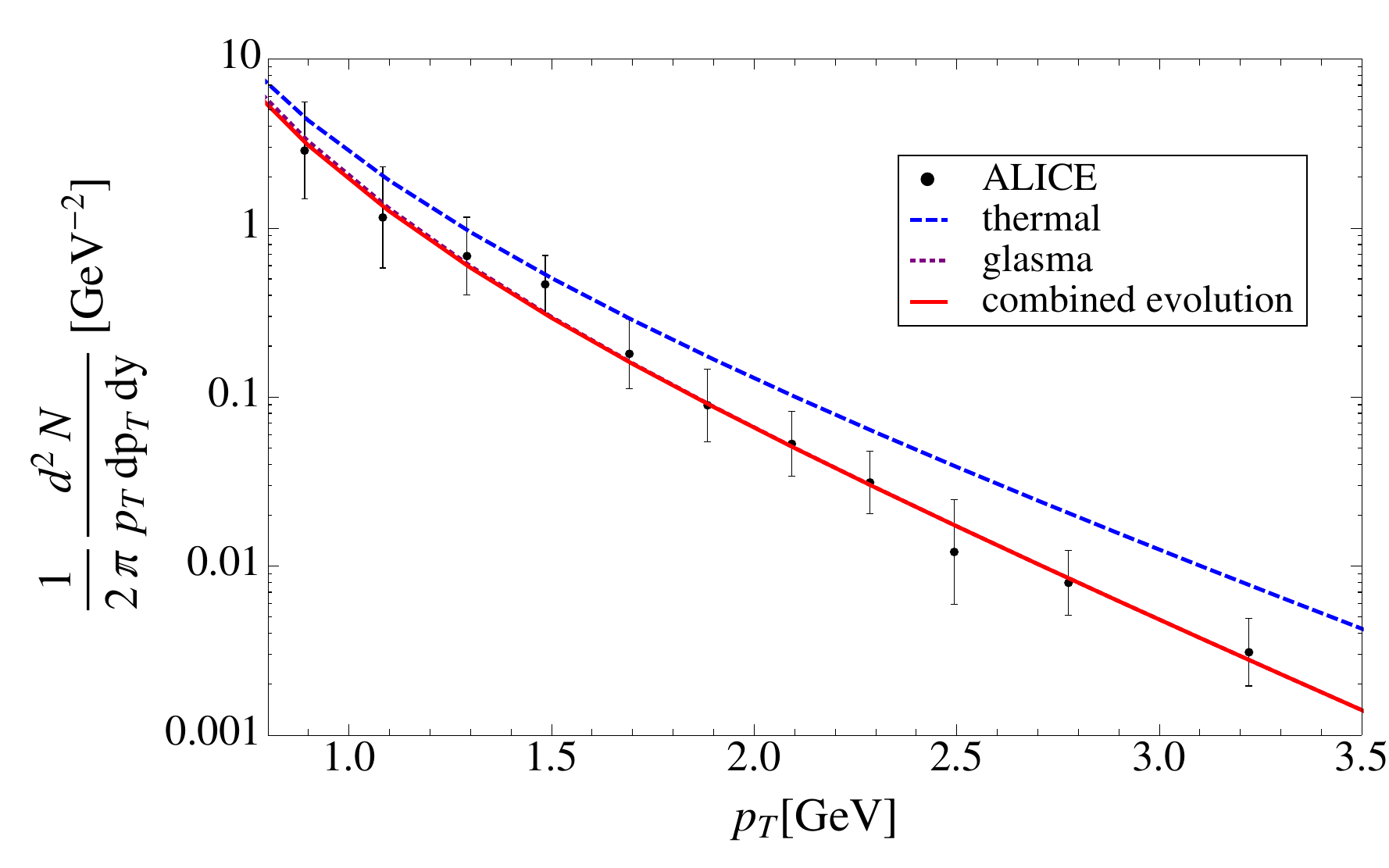}
\caption{The photon yield from the TQGP (dashed), the Glasma ($\kappa=3$, $\delta=1/3$) (dotted) and early time Glasma combined with later time thermal evolution (solid) compared to experimental data from the ALICE collaboration \cite{Wilde:2012wc}. \label{fig:photonYield-1over3}}
\end{center} 
\end{figure}

\begin{figure}[h]
\begin{center}
\includegraphics[width=12cm]{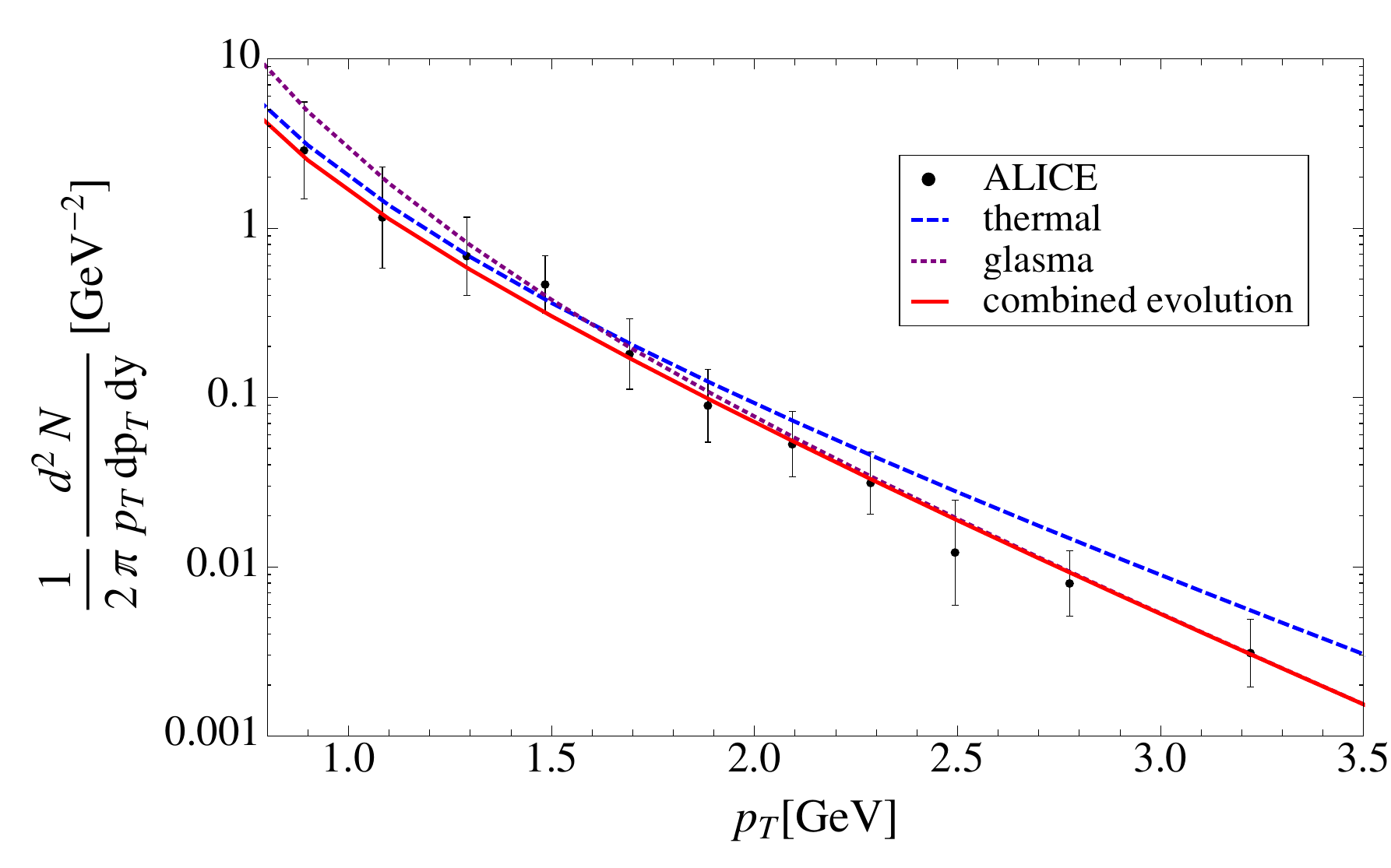}
\caption{The photon yield from the TQGP (dashed), the Glasma ($\kappa=3$, $\delta=1/4$) (dotted) and early time Glasma combined with later time thermal evolution (solid) compared to experimental data from the ALICE collaboration \cite{Wilde:2012wc}. \label{fig:photonYield}}
\end{center} 
\end{figure}

In Figs.\,\ref{fig:photonYield-1over3} and \ref{fig:photonYield} we show the time integrated photon yields for the thermal case
\begin{equation}
    \frac{1}{2\pi}{{dN_{\rm th}} \over {dy p_T dp_T}} = K \pi R^2 \int_{t_0}^\infty t dt \frac{dN_{\rm th}}{d^4x d^2p_Tdy}
\end{equation}
and the case of an initial Glasma stage
\begin{equation}
     \frac{1}{2\pi}{{dN_{\rm Glasma}} \over {dy p_T dp_T}} = K \pi R^2 \left(\int_{t_0}^{t_{\rm th}} t dt \frac{dN_{\rm Glasma}}{d^4x d^2p_Tdy}
+ \int_{t_{\rm th}}^\infty t dt \frac{dN_{\rm th}}{d^4x d^2p_Tdy}\right)
\end{equation}
assuming $R=7\,{\rm fm}$ and using a $K$-factor of 7 (for $\delta=1/3$) or 5 (for $\delta=1/4$), which accounts for logarithmic uncertainties in the rate, neglected collinear enhancement of bremsstrahlung, neglected non-collinear processes induced by jets \cite{Turbide:2007mi}, next-to-leading order (NLO) effects and photons from hadronic sources. In the regarded momentum range, these processes should not significantly modify the shape of the spectrum.
We also include a curve assuming Glasma emission for the entire lifetime of the system.

The photon distribution from the Glasma is clearly steeper than it is in the thermal case. Because we are using the thermal rates for times greater than the thermalization time in the combined evolution, low momentum photons that are emitted at later times, closely follow the thermal distribution. Higher momentum photons ($p_T \gtrsim 1\,{\rm GeV}$ for $\delta=1/3$ and $p_T \gtrsim 1.6\,{\rm GeV}$ for $\delta=1/4$) on the other hand dominantly emerge from the Glasma.

\begin{table}
\begin{center}
    \begin{tabular}{|c|c|c|c|c|}
    \hline
    $\langle t \rangle\,[{\rm fm}/c]$ & thermal & Glasma+th, $\delta=1/3$ & Glasma+th, $\delta=1/4$\\\hline
    $p_T=1\,{\rm GeV}$    & 12         &  16.8 &  15 \\\hline
    $p_T=2\,{\rm GeV}$    & 2          &  2.8  &  3  \\\hline
    $p_T=3\,{\rm GeV}$    & 1          &  1.3  &  1.5  \\\hline
    \end{tabular}
\end{center}
\caption{Mean photon emission time in a thermal system compared to the Glasma case with isotropic pressures ($\delta=1/3$) and a pressure anisotropy ($\delta=1/4$). In the Glasma case we switch to thermal emission after the thermalization time $t_{\rm th}$.\label{tab:meanTimes} }
\end{table}

For a more quantitative analysis we compute typical emission times for photons of different energies according to
\begin{equation}
 \langle t \rangle_{\rm th} = \frac{  \int_0^\infty t^2 dt\, dR_{\rm th}(t)/dy~p_Tdp_T }{  \int_0^\infty t dt\, dR_{\rm th}(t)/dy~p_Tdp_T}\,,
\end{equation}
and
\begin{equation}
 \langle t \rangle_{\rm Glasma+th} = \frac{  \int_0^{t_{\rm th}} t^2 dt\, dR_{\rm Glasma}(t)/dy~p_Tdp_T +  \int_{t_{\rm th}}^\infty t^2 dt\, dR_{\rm th}(t)/dy~p_Tdp_T  }{  \int_0^{t_{\rm th}} t dt\, dR_{\rm Glasma}(t)/dy~p_Tdp_T +\int_{t_{\rm th}}^\infty t dt\, dR_{\rm th}(t)/dy~p_Tdp_T }\,,
\end{equation} 
where $R_{\rm Glasma/th} = dN_{\rm Glasma/th}/d^4x$.

Results for three different photon energies are shown in Table \ref{tab:meanTimes}. We find that photon emission is shifted to significantly later times in the Glasma compared to the thermal system.

For $\delta=1/4$, the thermalization time is $t_{\rm th} \approx 8.6\,{\rm fm}/c$, for $\delta=1/3$ we find $t_{\rm th} \approx 9.7\,{\rm fm}/c$. 
Mean emission times that are significantly larger (as is the case for e.g. $1\,{\rm GeV}$ photons) indicate that a large fraction of photons are produced in a thermalized system, as we expected from the photon distribution shown in Fig.\,\ref{fig:photonYield}.

The slower expansion for $\delta=1/4$ causes emission times to increase, except at the lowest photon energies, where $\langle t \rangle$ is smaller than for $\delta=1/3$. This is because for $\delta=1/4$ thermalization is reached at a slightly earlier time than for $\delta=1/3$. Furthermore, because of the anisotropy, the rates make a small jump down at the thermalization time. At higher photon energies, the mean emission times are significantly smaller than the thermalization time and are not affected by this.

The later photon emission times for a system that undergoes evolution through a Glasma phase can have an important effect on the photon elliptic flow. The later the photons are produced the more flow the system will have built up by the time of emission. This effect could help explain the large experimentally observed photon elliptic flow.

\section{Conclusions}

We have shown in a simple model how photons might be emitted during the Glasma phase of ultra-relativistic heady ion collisions.  Our computations explicitly build
in the geometric scaling of photon distributions produced in such collisions.  Our computations give a reasonable agreement with the shape of the photon spectrum,
a feature which cannot be extracted by simple scaling arguments.  The time for emission of these photons in a combined Glasma+TQGP computation is significantly longer than
is the case for a pure TQGP computation.  Whether or not these increased times are sufficient to produce the observed flow seen in experiment can be determined by realistic
3+1 dimensional computations.  This paper outlines some of the ingredients which may be needed in such a computation.

\section*{Acknowledgments}

 The research of L.M. and B.P.S. is supported under DOE Contract No. DE-AC02-98CH10886.  Larry McLerran would like to thank Klaus Reygers and Johanna Stachel
 who organized the EMMI Rapid Reaction Task Force Meeting ``Direct Photon Flow Problem''.    He gratefully acknowledges the hospitality of EMMI.
 This work was initiated during this meeting, and was motivated by
 conversations specifically with Charles Gale and Ulrich Heinz, and more generally as a result of dialog with participants in the meeting.

\end{document}